\documentclass[psfig,dvips,useAMS,usenatbib]{mn2e}
\usepackage{times}
\usepackage{epsfig}
\usepackage{amsmath}
\usepackage{natbib}
\usepackage{longtable}
\usepackage{psfrag}
\usepackage{enumitem}
\def\mpc{h^{-1}{\rm{Mpc}}}
\def\kms {\rm{km~s^{-1}}}
\def\apj {ApJ}

\def\apjs {ApJS}
\def\aj {AJ}

\def\aap {A\&A}
\def\mnras {MNRAS} 

\def\sovast {Sov. Astron.}
\def\araa {Ann. Rev Astron. Astrophys.}
\def\nat {Nature}

\def\mpc{{\  h^{-1} \rm Mpc}}
\def\kpc{{\ h^{-1} \ \rm kpc}}
\def\kms{{\rm km \ s^{-1}}}

\title[Galaxy triplets in SDSS Data Release 7 - III]{Galaxy triplets in Sloan Digital Sky Survey Data Release 7 - III. Analysis of Configuration and Dynamics }

\author[Duplancic et al.]
{Fernanda Duplancic$^{1,2}$\thanks{E-mail: fduplancic@unsj-cuim.edu.ar}, Sol Alonso$^{1,2}$, Diego G. Lambas$^{4,2}$ \& Ana Laura O'Mill$^{3}$ 
\\
    $^1$ Departamento de Geof\'isica y Astronom\'ia, Facultad de Ciencias Exactas, F\'isicas y Naturales, Universidad Nacional de San Juan,\\ Av. Ignacio de la Roza 590 (O), J5402DCS - San Juan, Argentina \\
  $^2$Consejo Nacional de Investigaciones Cient\'\i ficas y T\'ecnicas (CONICET),  \\ Avenida Rivadavia 
1917, C1033AAJ - Buenos Aires, Argentina\\
 $^3$ Departamento de Astronomia, Instituto de Astronomia, Geof\'\i sica
e Ci\^encias Atmosf\'ericas da USP,\\ Rua do Mat\~ao 1226, Cidade
Universit\'aria, 05508-090 - S\~ao Paulo, Brazil.\\
$^4$ Instituto de  Astronom\'\i a Te\'orica y Experimental, IATE, Observatorio Astron\'omico, Universidad Nacional de C\'ordoba,\\ Laprida 854, X5000BGR - C\'ordoba, Argentina 
}

\date{\today}

\begin{document}
\pagerange{\pageref{firstpage}--\pageref{lastpage}}

\maketitle

\label{firstpage}

\begin{abstract}
We analyse the spatial configuration and the dynamical properties of a sample of 92 galaxy triplets obtained from the SDSS-DR7 (SDSS-triplets) restricted to have members with spectroscopic redshifts in the range $0.01\le z \le 0.14 $ and absolute r-band luminosities brighter than $M_r = -20.5$. 
The configuration analysis was performed through Agekyan \& Anosova map (AA-map). We estimated dynamical parameters, namely the radius of the system, the velocity dispersion, a dimensionless crossing-time and the virial mass. We compared our results with those obtained for a sample of triplets from the catalogue ``Isolated Triplets of Galaxies" (K-triplets) and a sample of Compact Groups. We have also studied a mock catalogue in order to compare real and projected configurations, and to estimate the three dimensional dynamical parameters of the triple systems.
We found that the SDSS-triplets prefer alignment configurations while K-triplets present an uniform distribution in the AA-map. From the dynamical analysis we conclude that the SDSS-triplets, K-triplets and Compact Groups present a similar behaviour comprising compact systems with low crossing-time values, with velocity dispersions and virial masses similar to  those of low mass loose groups. Moreover, we found that observed and simulated triplets present similar dynamical parameters. We also performed an analysis of the dark matter content of galaxy triplets finding that member galaxies of mock triplets belong to the same dark matter halo, showing a dynamical co-evolution of the system.
These results suggest that the configuration and dynamics of triple systems favour galaxy interactions and mergers.
\end{abstract}



\section{Introduction}

In the standard model of structure formation, the largest systems are formed from the accretion of less massive objects. Therefore, clusters and superclusters of galaxies are not only the largest gravitationally bound structures but are also the most recently formed. According to this hierarchical scenario, galaxies are preferentially clustered in configurations ranging from galaxy pairs to clusters that could contain hundreds of members. Between these two extremes lies a wide range of small galaxy systems such as triplets of galaxies, compact and loose groups. Moreover small galaxy groups are the most common galaxy associations containing about 50\% of all galaxies in the universe \citep{HG82}. In this hierarchical scenario of structure formation, the study of small groups as well as the properties of the galaxies that inhabit these systems, is essential to understand the relations between galaxy properties and the large-scale structure. 

The configuration of a triple system, defined as the shape of the triangle formed by their members, could be the result of the dynamical evolution of the group. Numerical simulations of three bodies-systems composed by particles with comparable masses show that during most of its lifetime the systems present a predominant type of configuration, named hierarchical, formed by a close binary and a remote third body \citep[e.g.,][]{Valtonen1991,Chernin94}. This scenario is significantly modified when the triplet is considered immersed in a dark matter halo that makes the dynamics of the whole system more stochastic, with no observed formation of close binaries within three crossing times \citep{Kiseleva2000}. Nevertheless \citet{Aceves2001} performed more consistent simulations of galaxy triplets by modelling the galaxies in the system as Plummer spheres and found, that a large primordial common halo of dark matter is not required to avoid the existence of hierarchical triplets.

From the observational point of view, there are two well studied samples, the catalogue of \citet{karachen88} (K-triplets) and the catalogue described in \citet{Wide} (W-triplets). The configuration of a sample of physical compact system derived from the K-triplets catalogue was studied by several authors \citep{Zheng93,Chernin94,Aceves2001,Melnyk2006}. In these works it is stated that the configuration of these systems is consistent with a random uniform distribution with no excess of any particular configuration type. \citet{Chernin94} demonstrate that these results is not due to projection effects implementing two methods, chance coincidence and inverse matrix, both based on computer generation of triplet sets an their transformation to projected systems.  \citet{Chernin94} found a weak excess of systems with hierarchical configuration for the sample of W-triplets concluding that this result is due to projection effects. Moreover, \citet{Aceves2001} found for the W-triplets an absence of equilateral-triangle type configuration. Nevertheless these results could be biased due to the smallness of the observational samples of triple systems. 

The foregoing suggest that the dynamics of observed triplets is inconsistent with the classic three-body problem and advocates for self-consistent galaxy models to study the dynamics of triplets and the role of a primordial common dark matter halo in a cosmological scenario.

In \citet{OMill2012} (hereafter paper I) we present a method to identify triplets of galaxies in the Seventh Data Release of the Sloan Digital Sky Survey. In \citet{Duplancic2013} (hereafter paper II) we analyse the properties of galaxies in a sample of isolated triplets of bright galaxies and found that these systems could be considered as an analogous of compact groups with a lower number of members. It is important to highlight that the spectroscopic sample analysed in paper II represents a statistically homogeneous sample with the largest number of objects analysed so far in the literature. This has allowed for a reliable study of the properties of triple systems and their member galaxies. Continuing with this series of papers, in the present work we focus on the characterization of the configurations and the dynamical properties of triplets of galaxies. 

This paper is organized as follows: In section \ref{data} we describe the different samples used in this work. Section \ref{conf} presents a description and analysis of the triplets configuration. The dynamical parameters of these systems are described in section \ref{dyn}  together with a comparison of these parameters with other galaxy systems samples. Finally in section \ref{con} we present the main results of this work.

Throughout this paper we adopt a cosmological model characterised by the parameters $\Omega_m=0.25$, $\Omega_{\Lambda}=0.75$ and $H_0=70~h~{\rm km~s^{-1}~Mpc^{-1}}$.

\section{Samples}
\label{data}

\subsection{SDSS-triplets}

In this work we analyse a samples of galaxy triplets 
derived from the Data Release 7 of Sloan Digital Sky Survey  \citep[SDSS-DR7,][]{dr7}.
SDSS \citep{sdss} has mapped more than one-quarter of the entire sky, 
performing  photometry and spectroscopy for galaxies, quasars and 
stars. SDSS-DR7 is the seventh major data release, corresponding to the
completion of the survey SDSS-II. It comprises $11663$ sq. deg.
of imaging data, with an increment of $\sim2000$ sq. deg., over the 
previous data release, mostly in regions of low Galactic latitude.
SDSS-DR7 provides imaging data for 357 million 
distinct objects in five bands, \textit{ugriz}, as well as
spectroscopy  over $\simeq \pi$ steradians in the North Galactic 
cap and $250$ square degrees in the South Galactic cap. 
The average wavelengths corresponding to the five broad bands 
 are $3551$, $4686$, $6165$, $7481$, and $8931$ \AA{} \citep{fuku96,hogg01,smit02}. 
For details regarding the SDSS camera see \citet{gunn98}; for astrometric 
calibrations see \citet{pier03}. 
The survey has spectroscopy over 9380 sq. deg.; the spectroscopy is now 
complete over a large contiguous area of the Northern Galactic Cap, closing the gap 
that was present in previous data releases. The spectroscopic data used in this work were derived from the Main Galaxy Sample of the SDSS-DR7 \citep[MGS;][]{mgs}. k-corrections for this sample were calculated bandshifted to $z=0.1$, using the software \texttt{k-correct\_v4.2} of \citet{kcorrect}. For this data, absolute magnitudes were calculated from Petrosian apparent magnitudes converted to the AB system.

 Following paper I and paper II, in this work we use a sample of triplets composed by three spectroscopic galaxies, brighter than $M_r=-20.5$, that are close in projected separation ($r_p <200\kpc$ whit respect to the centre of the system defined as the brightest galaxy of the triplet) and have a radial velocity difference $\Delta V_\mathrm{spec}<700 \kms$. In order to select isolated systems, we impose the condition that there are no bright ($M_r \leq -20.5$) neighbours within $0.5 \mpc$ from the triplet centre and with the same restriction on $\Delta V$ used to identify triplet members. For these isolated systems, we calculated the distance to the closest neighbour group/cluster ($d_c$) as described in paper II finding that 98\% of the triplets present a distance to the nearest cluster larger than $1.5\mpc$. Moreover, 95 \% of the systems are at distances greater than $ 3\mpc $ from these massive objects implying that the isolation criterion is effective in selecting systems far away from clusters of galaxies.

Under these considerations, the final sample of galaxy triplets contains 92 isolated systems with 276 spectroscopic galaxies  brighter than $M_r=-20.5$, in the redshift range $0.01\leq z\leq0.14$.

\subsection{K-triplets}

As a comparison sample, we also study the configuration and dynamics of triplets from the catalogue ``Isolated Triplets of Galaxies" of \citet{karachen88}. This catalogue contains the complete data for radial velocities of all components of triplets of galaxies from \citet{K79} that lists 84 triplets of galaxies in the northern sky (declination $\delta \geq -3^\circ$) obtained from the data of the Palomar Sky Survey and the catalogue of galaxies of \citet{Zwicky1968}.

The triplet identification criteria used in the construction of this sample considers galaxies with visual magnitude $m<$15.7. Each galaxy with angular diameter $a_1$ is examined as a possible member of a triplet in relation to significant neighbours, defined as those galaxies with diameters in the range [0.5$a_1$-2$a_1$]. These limits were introduced empirically in order to avoid contamination by foreground and background galaxies. A given set of three galaxies with diameters $a_1$, $a_2$ and $a_3$, which satisfy the above conditions, is considered as an isolated triplet if the distance to the nearest significant neighbour is more than 3 times the average distance between the galaxies in the triplet.  As significant neighbours the authors consider galaxies whose angular diameter differs by a factor lower than two times the average diameter of triplet members. This isolation criterion, chosen empirically, corresponds to density contrast of the triplet, relative to the background, greater than 10. 

The criteria described previously selected triplet member galaxies regardless their radial velocity, therefore, a system may be isolated only in projection. For these reason \citet{karachen88} compile the radial velocity monitoring data of these systems  and found that only 64\% of the triplets could be considered as physical in the sense that all its members satisfy the condition $\Delta V_{ij}<500 \ \kms$, were $\Delta V_{ij}$ is the radial velocity difference between galaxies $i$ and $j$.  Nevertheless \citet{Elyiv2009} found that compact triplet, with mean harmonic radius $R_h<200 \rm kpc$, are more isolated and have greater probability of being real physical systems. In contrast, wide triplets present a lower degree of isolation and, therefore, have a higher probability of being chance configurations. 

For these reasons in this work we will analyse only systems from \citet{karachen88} that have isolation indicator equal to (+++) in the catalogue. Also we imposed the restriction $\Delta V_{rms}<1000 \kms$ on the rms value of radial velocity difference of triplets member galaxies. This restriction was previously adopted by \citet{HT2011} in order to define physical triple systems in this sample. Only 37 triplets (111 galaxies), of the original catalogue of 84 systems, satisfy the above conditions and we will call them hereafter K-triplets. These systems belong to the local universe with redshifts in the range $0<z<0.05$ (with a mean value at $z \sim 0.02$), therefore non of them were found in the sample of SDSS-triplets.

\subsection{Mock Triplets}
In order to analyse real-space dynamical and configuration properties of triple systems, we used a mock catalogue derived from the semi-analytic models of \citet{croton} for the Millennium Simulation \citep{Springel2005} outputs. The spatial resolution of this simulation is suitable for the implementation of the triple system algorithm described in paper I. Also the semi-analytical model of \citet{croton} is consistent with observations of low-redshift galaxies, including the luminosity function, the global star formation history, the color-magnitude diagram and the Tully-Fisher relation. The mock catalogue comprises 514914 galaxies brighter than $r=17.77$, where $r$ correspond to the $r$-band magnitude of the SDSS, and contains relevant information as right ascension, declination, apparent magnitude in $r$-band, halo mass, redshift and peculiar velocity. It is important to highlight that we used this mock catalogue in paper I in order to estimate the completeness and purity of the galaxy triplets catalogues obtained through the implementation of our algorithm. 

We selected triplets in the mock catalogue considering three galaxies brighter than $M_r=-20.5$ with $d<200 \kpc$, were $d$ is the real distance from the centre, considered as the brightest galaxy of the triplet, and with no significant neighbours within $0.5\mpc$. The systems selected under these restrictions belong to the same dark matter halo. We take into account the resolution limit of the Millennium simulation and discard those systems with distances between member galaxies lower than 5$\rm kpc$. Under these restrictions we identified 159 triplets (477 galaxies) in the redshift range $0.01<z<0.14$.

It is worth to notice that the statistical analysis of configuration and dynamics developed in this paper will not be strongly affected by the presence of orphan galaxies, because the triplet members are bright ($M_r\leq-20.5$), whereas the orphans  primarily influence the fainter galaxy population \citep{Hearin2013}.

\subsection{Compact Groups}

We have compared the dynamical properties of triples with a sample of compact groups derived from the catalogue of \citet{McConnachie2009} who identified compact groups of galaxies in the photometric data of the SDSS-DR6, through the implementation of the `Hickson criteria' \citep{Hickson1982}. These authors provide two catalogues: Catalogue A comprising 2297 compact groups with member galaxies with petrosian $r$-band magnitude in the range $14.5\le r \le 18.0$ and Catalogue B including 74791 compact groups with galaxies in the broader magnitude range $14.5\le r \le 21.0$. Due to an independent visual inspection by the authors of all galaxy members in Catalogue A, the contamination of this sample due to gross photometric errors is negligible. Compact groups in Catalogue A have $\Delta v\le 1000 \kms$ where $\Delta v$ is a measure of the maximum line-of-sight velocity difference between group members for groups with more than two members with spectroscopic redshift information. Catalogue B includes many more groups than Catalogue A but has the disadvantage of a larger contamination due to poor photometric classification.

We will use Catalogue A as a comparison sample for this work. Nevertheless, this catalogue was constructed from the Sixth Data Release of the SDSS. Therefore we have added to Catalogue A spectroscopic redshift information from SDSS-DR7, and selected  systems that have all their galaxies with spectroscopic measurements and verify $\Delta v\le 1000 \kms$ taking into account the SDSS-DR7 redshift information.

We have cross-correlated our triplet sample with this compact group catalogue in order to exclude common systems finding only one coincidence. This common system consists on three triplet galaxies plus one galaxy fainter than $M_r=-20.5$. We have therefore removed this compact group from the original sample. Under these restrictions, we have selected 41 compact groups (168 galaxies) in the redshift range $0.01\leq z\leq 0.14$.

\section{Triple systems configuration}
\label{conf}
\subsection{The AA-map}
The configuration of a triple system is an important property in the study of its dynamical evolution. \citet{AA} suggested an elegant method to analyse the geometric configurations of a 3-body system. They propose a particular map (AA-map), where the shape of any triangle, despite of its proper size, could be represented by the position of a single point. In order to represent on this map a triplet with components A, B and C, it is necessary to perform a series of transformations such that the plane coordinates $(x_A,y_A)$ and $(x_B,y_B)$ of the two  particles at the extremes of the triangle largest side, are $(-1/2,0)$, $(1/2,0)$, respectively. The triplet configuration is defined by the coordinates $(x_C, y_C)$ of the third particle that should verify the following conditions: 

\setlength{\leftmargini}{3cm}
\begin{list}{•}{\addtolength{\itemsep}{1mm}}
\item $ x_C\geq 0 $ 
\item $y_C\geq 0$ 
\item $(x_C+0.5)^2+y_C^2\leq 1$
\end{list}

\begin{figure}
\centering
  \includegraphics[width=6.cm,height=6.cm]{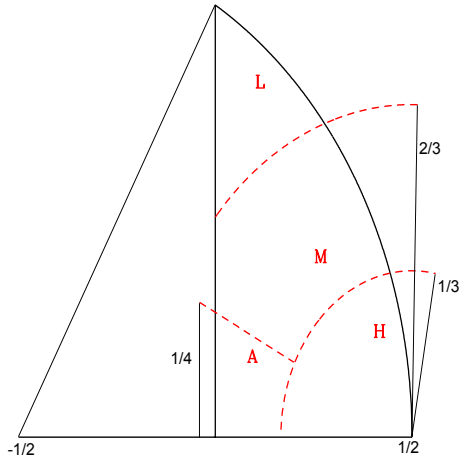}
\caption{AA-map with the different configuration areas: Lagrangian (L), Hierarchical (H), Alignment (A) and Intermediate (M).} 
\label{AAplot}
\end{figure}

Figure \ref{AAplot} shows a diagram of the AA-map. Through a geometrical analysis of configurations in this map, a triangle that has approximately equal sides will be represented by a point in the upper corner, a triangle with a much smaller side than the other two will be located in the lower right corner and linear configurations will be at the lower part of the map. According to these geometrical considerations, 4 areas could be defined in the AA-map:

\begin{itemize}[leftmargin=7mm,rightmargin=5mm]
\item \textbf{ L:}  Lagrangian region, represented by the points that verify $(x-0.5)^2+y^2\geq 2/3$. Systems whose configurations are similar to an equilateral triangle belong to this area. 
\item \textbf{ H:} Hierarchical region, defined by the points that verify $(x-0.5)^2+y^2\geq 1/3$. In this area inhabit the systems formed by a close binary and a remote third body. 
\item \textbf{ A:} Alignment region, represented by the points verifying $ y \leq -0.5 \ x +1/4$, is the area of systems with three aligned galaxies. 
\item \textbf{ M:} An intermediate region located between the three areas defined previously. 
\end{itemize}

This map has been used by several authors to characterize the configuration of triplets of galaxies in different catalogues and to analyse its relation with diverse dynamic properties of the systems \citep{Chernin94,Aceves2001}.

\subsection{Real-space versus projected configurations}
\label{3D2Dconfig}
The configuration of a triplet, defined through the AA-map, is based on the projected configuration and not on the study of the actual three-dimensional configuration of the system. The influence of projection effects on the AA-map were analysed by \citet{Chernin94}. These authors found that a set of simulated configurations of triplets with uniform density distribution on the real AA-map becomes highly non-uniform after projection. Also the projection effect increases the number of triplets in the hierarchical area.

Different authors  assess the projection effect on the AA-map through the analysis of three-body simulations. In this paper we consider a different approach that involves the analysis in a cosmological setting through the use of a SDSS-DR7 synthetic catalogue.

\begin{figure}
\centering
  \includegraphics[width=8.5cm,height=9cm]{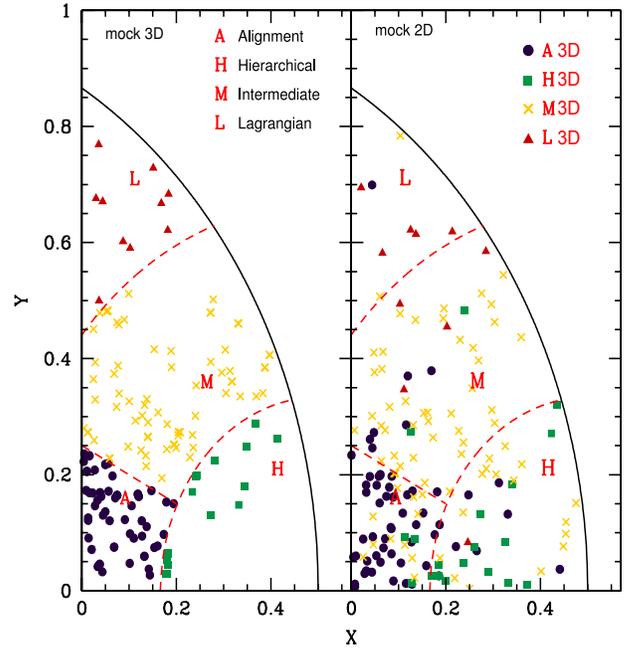}
\caption{AA-map for the triplets derived from the mock catalogue. \textit{Left:} Real 3D configurations. \textit{Right:} Projected 2D configurations. } 
\label{mock3D2D}
\end{figure}

For the mock triplets described in section \ref{data} we calculate the real three-dimensional configurations (3D) as well as the projected (2D) configurations on the AA-map. Figure \ref{mock3D2D} shows the real and projected AA-map for the triplets in the mock catalogue. We used different colors and symbols in order to discriminate 3D system configuration in the 2D AA-map, the purpose of this analysis is to estimate the percentage of triplets whose projected configurations match their real ones. The resolution limit of the Millennium simulation affects mostly the hierarchical area giving rise to the lack of systems in the right corner of the AA-map. The A and M regions are also affected but in a negligible percentage.

\begin{table}
\center
\caption{Number of mock triplets in each area of the 3D and 2D AA-map and percentages of 3D2D and 2Dno3D systems.}
  \begin{tabular}{c c c c c}
  \hline
  \hline
 Configuration & 3D & 2D & 3D2D & 2Dno3D\\
  \hline
  \hline
A  & 59 & 59 & 73\% & 27\% \\  
H  & 19 & 39 & 74\% & 64\% \\        
M  & 71 & 53 & 56\% & 24\% \\
L  & 10 &  8 & 50\% & 37\% \\
 \hline
\hline
\label{conf2d3d}
\end{tabular}
\end{table}

Table \ref{conf2d3d} shows the number of systems en each area of the 3D AA-map and in the 2D AA-map, also the percentage of systems that retain their real configurations after projection (3D2D) and the percentage of triplets in each area that do not match their real 3D configurations (2Dno3D). In each area of the 2D AA-map more than 50\% of the triplets match their 3D configurations.  However, the contamination by 2Dno3D triplets is significant in the H zone of the 2D AA-map. This result is in agreement with \citet{Chernin94}, that found that the projection effect produces an excess of systems in the H area. Through the present analysis, we consider that most of this contamination (60\%) is due to triplets with M-type real configuration.

In general, approximately 64\% of the mock triplets retain they real configurations after projection.  Therefore we can conclude that the AA-map is an appropriate tool for the analysis of triple systems configurations, as it represents consistently the real configuration of the triplets in all the AA-map areas, except in H area where the contamination changes the relative distribution of this region.

\subsection{Configuration analysis}

\begin{figure}
\centering
  \includegraphics[width=6cm,height=10cm]{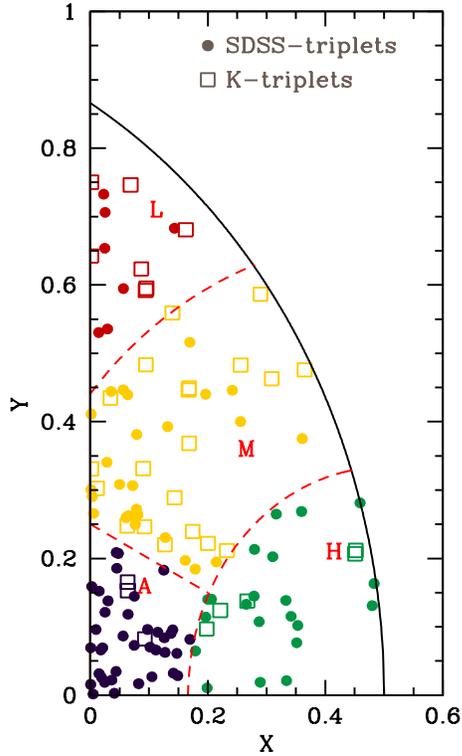}
\caption{AA-map of SDSS-triplets and K-triplets samples.} 
\label{AA_map}
\end{figure} 

In the previous section we used a mock catalogue and proved that the AA-map correctly represents the 3D configuration of triple systems. Therefore, it is interesting to analyse the configurations of the triplets of galaxies in the samples under study. 

Figure \ref{AA_map} shows the configuration of the SDSS-triplets and K-triplets. From this figure it can be appreciated that there is no excess of hierarchical systems, composed of two very close galaxies and a distant third member (right bottom of the AA-map) in both the SDSS-triplets and the K-triplets samples.

In order to correctly quantify the number of objects in each area of the AA-map, we calculate the relative number density of systems as $ \alpha_i=\rho_i\ / \rho $ where $ \rho_i $ is the number density of triplets in the area $ i $ (with $ i $ = A, H, M, L)  and $ \rho $ is the number density of systems in the entire AA-map.

\begin{figure}
\centering
  \includegraphics[width=7.5cm,height=7.5cm]{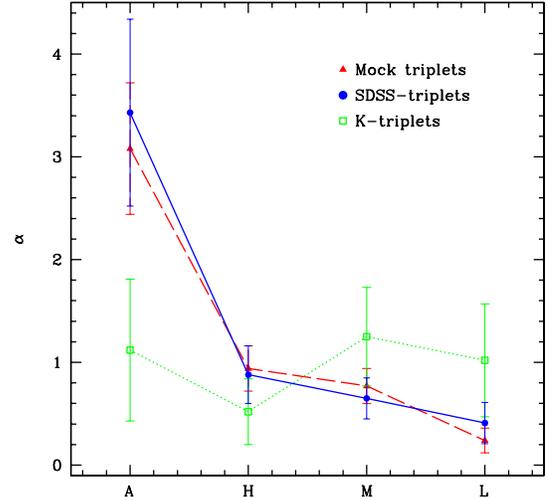}
\caption{Relative density of triplets ($\alpha$) as a function of the configuration type in the AA-map, for the 2D mock triplets, the SDSS-triplets and the K-triplets. Error bars correspond to the standard error.} 
\label{alphas}
\end{figure}

Table \ref{densities} shows the density in each area of the AA-map, for the SDSS-triplets and K-triplets. For comparison in this table the values of $\alpha $ for the 2D AA-map of triplets from the mock catalogue are also shown. Figure \ref{alphas} presents a graphical representation of this values as a function of the different configurations. From the analysis of the relative density in each area of the AA-map we conclude that most of the SDSS-triplets, as well as the triplets from the mock catalogue, present type A configuration. In the case of K-triplets there is no strong preference of these systems to have a particular configuration in the AA-map, in agreement with the results found in previous works \citep[e.g.][]{Aceves2001,Zheng93}. 

In order to explore selection effects, we constructed a subsample of SDSS-triples that verify the selection criteria and isolation of \citet{K79}. As a measure of galaxy size, we considered the Petrosian radius $r_{90}$ (defined as the radius enclosing $90\%$ of the Petrosian flux of the galaxy in the r-band) and for the selection of significant neighbours we applied a velocity difference cut of 1000 $\kms$. From the original sample of 92 SDSS-triplets, 44 systems verify the selection criteria of \citet{K79}, nevertheless these systems still prefer type A configurations.

\begin{table}
\begin{minipage}{90mm}
 \caption{ Relative density of triplets in each area of the AA-map. Uncertainties correspond to the standard error.}
  \begin{tabular}{l c c c c}
  \hline
  \hline
 Sample &  $\alpha_A $ & $\alpha_H$ & $\alpha_M$ & $\alpha_L$ \\
  \hline
  \hline
Mock triplets & 3.08$\pm$0.64 & 0.94$\pm$0.22 & 0.77$\pm$0.17 & 0.27$\pm$0.12\\  
SDSS-triplets  & 3.43$\pm$0.91  &  0.88$\pm$0.28 & 0.65$\pm$0.20 & 0.41$\pm$0.20  \\        
K-triplets  & 1.12$\pm$0.69 & 0.52$\pm$0.32 & 1.25$\pm$0.48 & 1.02$\pm$0.55\\
 \hline
\hline
\label{densities}
\end{tabular}
\end{minipage}
\end{table}

The AA-map describes the configuration of a triple system regardless its compactness. In order to investigate if there is a configurations that favours more compact systems we consider the compactness parameter $$S= \dfrac{\sum_{i=1}^{N} r_{90}^2}{R^2}$$ 

Where $R$ is the minimum enclosing circle that contains the geometric centres of the galaxies in the system, $r_{90}$ is the Petrosian radius defined previously and N is the total number of members of the system. This quantity, defined in paper II, is a measure of the percentage of the system total area that is filled by the light of member galaxies.    

We calculated the compactness $S$ for SDSS-triplets and K-triplets samples, but not for mock triplets because we do not have the $r_{90}$ parameter for mock galaxies. For the K-triplets we only have the $r_{90}$ value of 24 galaxies that present photometry in the SDSS-DR7. Therefore, we used these galaxies to perform a linear fit of $r_{90}$ as a function of galaxy major axis, $\rm MajAxis$, (extracted from  the ``Isolated Triplets of galaxies'' catalogue). We used this fit to estimate the $r_{90}$ value for the K-triplets galaxies, which allows us to calculate the $S$ parameter for this sample.
  
\begin{figure}
\centering
  \includegraphics[width=8.5cm,height=4.5cm]{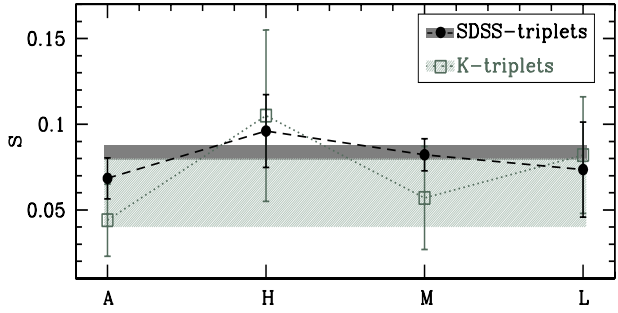}
\caption{Compactness $S$ median value for SDSS-triplets and K-triplets as a function of AA-map configuration type. The shaded regions represent the confidence interval of the median for each sample. Errors were calculated using bootstrap resampling techniques.} 
\label{S}
\end{figure}

Figure \ref{S} shows the S median values as a function of the triplet configuration, for the SDSS-triplets and K-triplets samples. The shaded regions represent the confidence interval of the median of $S$ for each sample. The errors were calculated using bootstrap resampling techniques \citep{boot}. From this figure it can be appreciated that there is no strong correlation between compactness and configuration type and that there is no difference between the compactness of SDSS-triplets and K-triplets. 

\section{Dynamical properties of triple systems}
\label{dyn}
\subsection{Dark matter in triple systems}
There is a current controversy regarding the distribution of dark matter in galaxy triplets. Several authors have performed simulations of triple systems requiring the presence of a primordial massive dark matter halo in order to correctly describe the dynamics of these systems \citep[e.g.][]{Valtonen1991,Chernin94,Kiseleva1993,Kiseleva2000}. Nevertheless \citet{Aceves2001} states that the presence of a dark matter halo would not be a strong requirement to describe the dynamical evolution of triplets.

In order to asses the dynamics of triplets we have analysed the mock triplet systems described in section \ref{data}. It is worth to notice that all the triplets identified in the mock catalogue have galaxy members that belong to the same dark matter halo. This result is a strong evidence of the dynamical co-evolution of the system. Furthermore, the presence of an underlying dark matter halo suggests that the system is dynamically evolved, giving time to the individual halos to merge.

\begin{figure}
\centering
  \includegraphics[width=8.5cm,height=8.5cm]{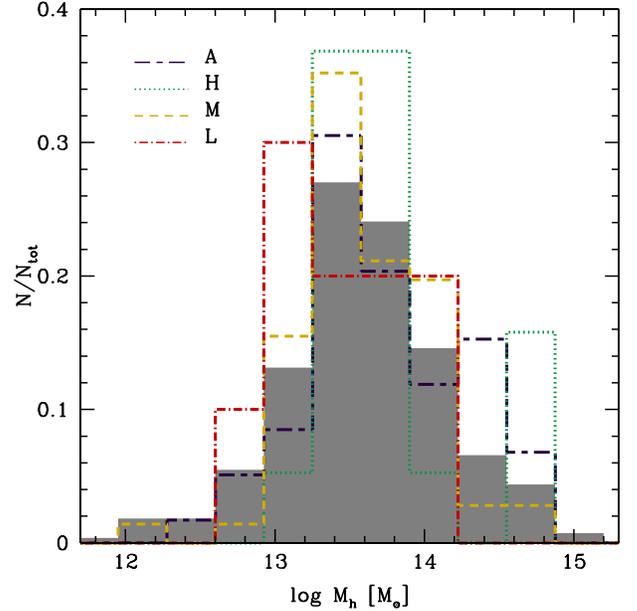}
\caption{Halo mass distributions of the mock triplets considering the different 3D AA-map configurations.} 
\label{halo_mass}
\end{figure}

The halo mass distribution of mock triples described in section \ref{data} is shown in figure \ref{halo_mass}. The typical halo mass is about $2 \rm x 10^{13} M_{\odot}$ which is comparable to the halo mass of compact groups according to \citet{McConnachie2008}. In this figure we also plot the distribution for the different 3D AA-map configuration and found no significant difference in the average value of the masses of the halos according to the system configuration. However, for massive halos ($M_h>1.5\rm x 10^{14} \ M_{\odot}$) the A and H distributions present a fraction of triplets of 25\% and 22\%, respectively, compared to the M and L systems which fractions are 6\% and 0\%, for halos of these masses. On the other hand there is a weak tendency of L systems to reside in less massive halos. In the latter case the number of systems is low so that the distribution may not be representative of the behaviour of L-type triplet population.

 \begin{figure}
\centering
  \includegraphics[width=8.cm,height=5.5cm]{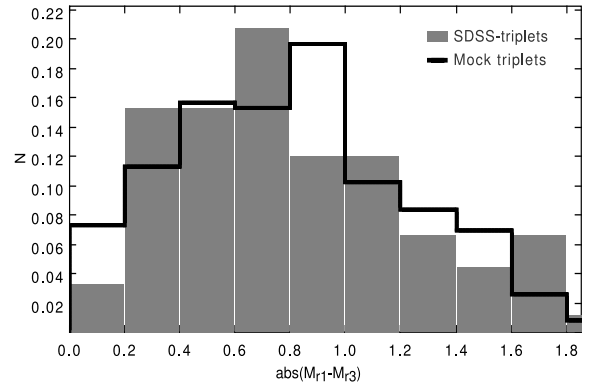}
\caption{Magnitude difference between the brightest and the faintest galaxy in the triplet, for SDSS-triplets and for mock triplets.} 
\label{lumdif}
\end{figure}

The relative luminosity  of the triplet member galaxies could be of importance for the dynamical interpretation of the system. Figure \ref{lumdif} shows the maximum magnitude difference between triplet members, i.e. the difference between the magnitude of the brightest and the faintest galaxy in the system, for SDSS-triplets and for mock triplets. From this figure it can be appreciated that the average maximum magnitude difference is 0.8 for SDSS-triplets and 0.9 for mock triplets. Moreover, the difference between galaxy members is lower than 2 magnitudes that is the limit adopted for \citet{Sales2005} to select satellite galaxies. Therefore, triplets comprise similar galaxies that reside in a single halo, these results suggest that the triplets have probably undergone recent merger events.

\subsection{Dynamical analysis}
  \begin{table*} 
\center
\caption{Median of the dynamical parameter for the samples analysed in this work.}
  \begin{tabular}{l c c c c}
  \hline
  \hline
 Sample &  $ R $ & $ \sigma_{\rm v} $ & $ H_0 t_c $ &$ \rm log (M_{vir}) $ \\
 		&   [$\kpc$]  & [$\rm km/s$] & & [$\rm M_{\odot}$] \\
  \hline
  \hline
Mock triplets &  42.7$\pm$1.3 & 126.9$\pm$13.2 & 0.019$\pm$0.002 &11.8$\pm$0.1\\
SDSS-triplets  &  67.1$\pm$2.8 & 118.6$\pm$8.9& 0.031$\pm$0.002 & 12.0$\pm$0.1 \\      
K-triplets  & 53.7$\pm$5.6 & 113.8$\pm$12.2 & 0.019$\pm$0.006 & 11.9$\pm$0.1\\
Compact Groups  & 81.3$\pm$5.5 & 199.3$\pm$16.6 & 0.021$\pm$0.003 &12.5$\pm$ 0.1\\ 
 \hline
\hline
 \label{dynamical}
\end{tabular}
\end{table*}

In order to characterize the system dynamics we consider the radius, the velocity dispersion, the crossing time and the virial mass of the systems.

For the crossing time ($t_c$) and virial mass ($M_{\rm vir}$) we use the following expressions \citep{Hickson1992}: 

$$t_c={4\over \pi} \ {R \over {\sqrt 3 \sigma_{\rm v}}} $$
$$M_{\rm vir}={{3 \pi} \over {2G}} R  \sigma_{\rm v}^2$$

Where $\sigma_{\rm v}^2=\langle V^2 \rangle-\langle V \rangle^2 $ is the velocity dispersion of the system, $V$ is the radial velocity of the galaxy and $\langle \rangle$ denotes the average over all the members of the system. Following paper II we have calculated the radius of the system $R$ as the minimum circle containing the centres of the group member galaxies, using the code developed by \citet{minicirc}. We have also calculated the dimensionless crossing time $H_0 \ t_c$, which is the ratio of the crossing time to the age of the universe and is a convenient measure of the dynamical state of the system.

In order to perform a comparative analysis, we have studied the dynamics of the SDSS-triplets, K-triples and Compact Groups samples. We have also estimated the dynamical parameters of the mock triplets in order to compare the dynamics of simulated and observed triple systems. Table \ref{dynamical} summarizes the main results. It can be appreciated that SDSS-triplets, K-triplets and Compact Groups present similar median dynamical parameters with low crossing time values. We also found that the velocity dispersion and virial masses of SDSS-triplets are similar to low mass loose groups \citep{MZ2005}.  It is worth to notice that the estimated virial mass of SDSS-triplets differ by more than one order of magnitude from the typical halo mass analysed in section \ref{halo_mass}. Nevertheless this result is in agreement with \citet{McConnachie2008} who found that the mass of compact groups, calculated through the virial theorem, is a gross underestimate of the halo mass. The authors state that galaxies are embedded in the central region of the dark matter halo and sample a smaller mass corresponding to the innermost region of the halo.

From table \ref{dynamical} it can also be appreciated that the dynamical parameters obtained for the mock triplets are similar to those found for the SDSS-triplets. This result demonstrates that triplets obtained from cosmological simulations and observed triple systems show a concordant dynamical behaviour.

\begin{figure}
  \includegraphics[width=8.5cm,height=9.5cm]{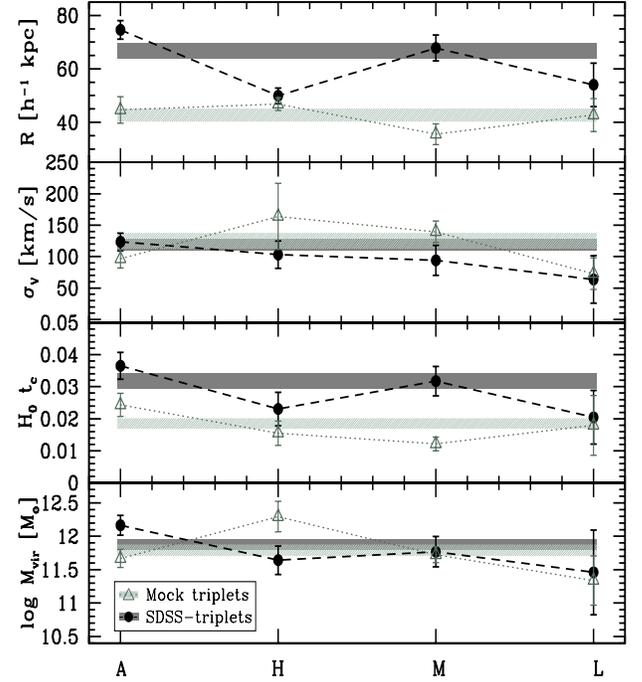}
\caption{Median of the dynamical parameters for the SDSS-triplets and mock triplets for the different AA-map configurations. The shaded region corresponds to the confidence interval for the median for the entire sample. Errors were calculated using bootstrap resampling techniques.} 
\label{dynamic_config}
\end{figure}

It is expected that dynamical effects are most pronounced in systems with low values of $H_0 \ t_c $. Therefore, the compact configuration could be a result of the dynamical evolution of the system. This scenario may favour the formation of elliptical galaxies through mergers, in agreement with the results found in paper II that SDSS-triplets present an important fraction of bulge-type, red galaxies, with low star formation activity indicators, similarly to galaxy members of compact groups. 

It is also important to consider that different configurations of the triplets may be reflecting different dynamical stages of the system. For this reason we have calculated the median of the dynamical parameters for the different AA-map configuration types for the SDSS-triplets and Mock triplets. In figure \ref{dynamic_config} we show the results obtained where the shaded regions represent the confidence interval of the median of the parameters for each sample. From this figure it can be appreciated that there are no strong trends of the dynamical parameters with respect to the different AA-map configurations for the SDSS-triplets nor for mock triplets. Nevertheless, despite the error bars, it can be appreciated that SDSS-triplets with A-type configuration present higher values of $R$ and $\sigma_{\rm v}$ which results in larger crossing time and greater values of $M_{\rm vir}$. In average the mock triplets are more compact than SDSS-triplets, therefore the crossing times are lower, this is more evident for type A and type M configurations where the median values of $R$ and $H_0 t_c$ of mock and SDSS-triplets  differ significantly. Nevertheless it is worth to notice that mock and real triplets are similar in all other aspects analysed in this work.

\section{Summary and Conclusions}
\label{con}
In this work we study the configuration and dynamical properties of a sample of triple galaxy systems derived from the Seventh Data Release of the Sloan Digital Sky Survey (SDSS-triplets). We used a mock catalogue in order to compare the configuration and dynamics of the observed and simulated galaxy triplets. In order to compare the results with those obtained for similar galaxy systems, we analysed a sample of galaxy triplets obtained from the catalogue ``Isolated Triplets of Galaxies" of \citet{karachen88} (K-triplets) and a sample of compact groups selected from the Catalogue A of \citet{McConnachie2009}. 

We can summarize the main results of this work as follows:
 
\begin{enumerate}[leftmargin=3mm,rightmargin=2mm,itemindent=6mm]
\item Our analysis of the real-space and projected configuration of triple systems allows to conclude that the AA-map  provides a suitable characterization of the true configuration of the triplets because about 64\% of the triplets retain their real configuration after projection. Nevertheless, in the H area of the 2D AA-map the contamination by 2Dno3D triplets is significant and changes the relative distribution of this region. In this sense, we conclude that there is a large contribution of this contamination (60\%) by triplets with M-type true configurations. 

\item For the SDSS-triplets we find that most of these systems present an alignment type configuration (A) while K-triplets show a more uniform distribution on the AA-map. The preference of A-type configuration persist even for a sample of SDSS-triplets that fulfil the selection  and isolation criteria of \citet{K79}. Nevertheless, there is no strong correlation between compactness and configuration, for SDSS-triplets nor for K-triplets.

\item The study of the mock triplets allow for an analysis of the dark matter content of galaxy triple systems. We find that the dark matter halos of triplets have masses similar to those of compact and loose galaxy groups. We also find that member galaxies of mock triplets reside in a common dark matter halo, suggesting that triple systems are dynamically evolved. 

\item We also study the maximum magnitude difference between galaxy members in both mock and SDSS-triplets. We found that triplets comprise similar galaxies that reside in a single halo, these results suggest that the triplets have probably undergone recent merger events.

\item We find that SDSS-triplets, K-triplets and Compact Groups present similar dynamical parameters. Also the dynamics of mock triplets are similar to those found for the SDSS-triplets, showing that simulated and observed systems present a concordant dynamical behaviour.

\item The low values of $H_0\ t_c$ found for SDSS-triplets suggest that dynamical effects are most pronounced in these systems. Moreover, the different configuration could be the result of the dynamical evolution of the triplets. Nevertheless we found no strong trends of the dynamical parameters with respect to the different AA-map configurations for the SDSS-triplets nor for mock triplets.

\end{enumerate}
We conclude from these results that, in SDSS-triplets, the formation of elliptical galaxies through mergers could be favoured by the configuration and dynamics found in this work. This is in agreement with the results given in \citet{Duplancic2013} who show that SDSS-triplets present an important fraction of bulge-type, red galaxies, with low star formation activity indicators, similarly to galaxy members of compact groups.


\section{Acknowledgements}
We thank the referee, Nelson Padilla, for providing us with helpful comments that improved this work. This work was supported in part by the Consejo Nacional de 
Investigaciones Cient\'ificas y T\'ecnicas de la Rep\'ublica Argentina 
(CONICET). ALO was supported by the Brazilian agencies FAPESP and CNPq. 
Funding for the SDSS and SDSS-II has been provided by the Alfred P. Sloan Foundation, 
the Participating Institutions, the National Science Foundation, the U.S. 
Department of Energy, the National Aeronautics and Space Administration, 
the Japanese Monbukagakusho, the Max Planck Society, and the Higher Education 
Funding Council for England. The SDSS Web Site is http://www.sdss.org/. 
The SDSS is managed by the Astrophysical Research Consortium for 
the Participating Institutions. The Participating Institutions are 
the American Museum of Natural History, Astrophysical Institute 
Potsdam, University of Basel, University of Cambridge, Case 
Western Reserve University, University of Chicago, Drexel 
University, Fermilab, the Institute for Advanced Study, the Japan 
Participation Group, Johns Hopkins University, the Joint Institute 
for Nuclear Astrophysics, the Kavli Institute for Particle 
Astrophysics and Cosmology, the Korean Scientist Group, the 
Chinese Academy of Sciences (LAMOST), Los Alamos National 
Laboratory, the Max-Planck-Institute for Astronomy (MPIA), the 
Max-Planck-Institute for Astrophysics (MPA), New Mexico State 
University, Ohio State University, University of Pittsburgh, 
University of Portsmouth, Princeton University, the United States 
Naval Observatory, and the University of Washington.



\end{document}